\pdfoutput=1

\documentclass[10pt,journal,twoside,compsoc]{IEEEtran}
\ifCLASSOPTIONcompsoc
  \usepackage[nocompress]{cite}
\else
  \usepackage{cite}
\fi
\ifCLASSINFOpdf
   \usepackage[pdftex]{graphicx}
\else
\fi

\usepackage{array}

\usepackage{amsmath,amssymb, amsbsy,bm}
\usepackage{commath}
\usepackage{xcolor}
\usepackage{booktabs}
\usepackage[caption=false,labelformat=simple]{subfig}

\usepackage{url}
\usepackage[normalem]{ulem}
\usepackage{pinlabel}
\usepackage[breaklinks, hidelinks,colorlinks=false]{hyperref}

\newcolumntype{L}[1]{>{\arraybackslash}p{#1}}
\newcolumntype{X}[1]{>{\raggedright\let\newline\\\arraybackslash\hspace{0pt}}m{#1}}

\newcommand\DSon{DSSon}

\newcommand\trend{\text{trend}}
\newcommand\AC{\text{AC}}

\newcommand\rf{\text{ref}}
\newcommand\up{+}
\newcommand\down{-}
\newcommand*{\s}[1]{\mathring{#1}}
\newcommand\td{t}

\newcommand\comp{\s{\kappa}}
\newcommand\dilate{\s{\Delta}}
\newcommand\thresh{\s{\epsilon}}
\newcommand\scalea{\s{\alpha}}
\newcommand\scaleb{\s{\beta}}
\newcommand\distort{\s{\phi}}
\newcommand\sont{\s{t}}
\newcommand\sonT{\s{T}}
\newcommand\decay{\s{\gamma}}
\newcommand\timbre{\s{H}}

\newcommand\gain{\s{g}}
\newcommand\snu{\s{\nu}}
\newcommand\ssigma{\s{\sigma}}
\newcommand\sJ{\s{J}}
\newcommand\sA{\s{A_0}}

\begin{document}

\title{Direct Segmented Sonification of\\ Characteristic Features of the Data Domain}
%
%
\author{Paul Vickers
and Robert H\"oldrich (IEEE member)
\thanks{P. Vickers is with the Department of Computer and Information Sciences, Northumbria University, Newcastle upon Tyne, UK e-mail: paul.vickers@northumbria.ac.uk.}
\thanks{R. H\"oldrich is with the Institute of Electronic Music and Acoustics, University of Music and Performing Arts Graz, Austria e-mail: robert.hoeldrich@kug.ac.at.}
\thanks{Manuscript received month dd, 2017; revised month dd, 2017.}}

\markboth{Pre-print}%
{Vickers \& H\"oldrich:  Direct Segmented Sonification}
%


\IEEEtitleabstractindextext{%
\begin{abstract}
Sonification and audification create auditory
displays of datasets. Audification translates data points into digital 
audio samples and the auditory display's duration is determined 
by the playback rate. Like audification,
auditory graphs maintain the temporal relationships of 
data while using parameter mappings (typically data-tofrequency) to represent
the ordinate values. Such direct approaches have the advantage of presenting
the data stream `as is' without the imposed
interpretations or accentuation of particular features found in 
indirect approaches. However, datasets can often be subdivided into
short non-overlapping variable length segments that each
encapsulate a discrete unit of domain-specific significant information
and current direct approaches cannot represent these. We present Direct Segmented
Sonification (\DSon) for 
highlighting the segments' data distributions as individual sonic events.
Using domain knowledge to segment data, \DSon\ 
presents segments as discrete auditory gestalts while retaining the
overall temporal regime and relationships of the dataset.
The method's structural decoupling from the sound stream's formation means playback
speed is independent of the individual sonic event durations, thereby offering highly flexible
time compression/stretching to allow zooming into or out of the data. Demonstrated by three models
applied to biomechanical data, \DSon\ displays high directness, letting the data `speak' for themselves.
\end{abstract}

}

\maketitle

%
\IEEEraisesectionheading{
\section{Introduction}}
\IEEEPARstart{S}{onification} 
and audification are representational techniques in which data sets, or selected features of data sets, are mapped to audio signals (see \emph{The Sonification Handbook} \cite{Hermann:2011} for an overview of the main techniques.)  As sound is a temporal phenomenon such auditory displays are especially well suited to time series data.

The most common sonification method is via parameter mapping in which the data values drive the parameters of an audio signal. For example, Parseihian et al \cite{Parseihian:2016} mapped target distance to pitch, timbre, and tempo in various combinations to assist with one-dimensional guidance tasks. Silva et al \cite{Silva:2016} mapped features of graphical objects to acoustic parameters to communicate visual information to visually impaired people. Using physical modeling sound synthesis, 
Roodaki et al. \cite{Roodaki:2017} mapped stylus pressure to the timbral parameters of an acoustic model to assist users with visual object tracking tasks.

In contrast, audification only involves transposing the frequencies of the data to the human-audible range and occasional filtering to remove unwanted linear distortions (and in rare cases dynamic range compression to remove very large level variations). Therefore, the process maintains a tighter relationship with the data than other auditory display processes which generally rely on mappings to effect the auditory display. These mappings can be low level (e.g., \cite{Parseihian:2016}) or more metaphorical (e.g., the use of melodic phrase structures to represent elements of a computer program \cite{Vickers:2005c}).

The directness  of a sonification  is a measure of the arbitrariness (in relation to the underlying data) of the mapping \cite{Vickers:2006a}. A  method exhibiting maximal directness will derive the sound directly from the data (e.g., through the use of direct data-to-sound translations). Low directness arises from more symbolic, metaphoric, or interpretative mappings. Thus, audification is a more direct form of auditory display, the audio being generated entirely by the data.  

The sonification method proposed here pursues directness as a design goal so that, as far as possible, the data are allowed to `speak' for themselves. In this way, any metaphors arise as  contingent properties of the sonification rather than being imposed by the designer. For example, the characteristic sound caused by accentuating data range excursions in \S\ref{sec:adv} below assumes its own sonic identity and metaphorical labels may be assigned by (and will vary depending on) the listener. In this way, sonification users may start identifying regions of interest in the data by describing the characteristic sounds they hear.

The proposed method follows a direct sonification strategy which conserves fundamental properties of (pure) audification, notably the compact temporal support and some aspects of the precise temporal structure of a data set. 

\subsection{Leveraging the Directness of Audification}
\label{sec:audgraphs}
The audification of a physical process strictly conserves the temporal regime of the source signal and so contains high-frequency components when rapid transients occur in the data. This is advantageous because such transients, which often correspond to points of interest in the data, are also significant features of the audio signal which the human auditory system relies on to identify real-world sounds. Hence, they can be a perceptually salient basis for auditory data exploration \cite{Holdrich:2015}.

When the data is sampled from a band-limited physical process the audification signal has a one-to-one relationship with the data.
In fact, the mapping is, in principle, bijective and fully reversible (at least while the data remains in the digital domain prior to any D/A conversion.)  However, even such direct representations can contain misleading features because of the band-limited interpolation of the reconstruction filter of the D/A converter leading to extreme data values being elevated in the audification. 

As H\"oldrich and Vogt \cite{Holdrich:2015} pointed out, the ideal audification signal has auditory gestalts within time and frequency ranges that are clearly perceptible to a listener. Take a data stream dominated by low frequencies with transients occurring within a range of 1 k data points and with an aperiodic interval of approximately 10 k data points. At a playback rate of 44.1 kHz roughly four of these events will occur each second which is comparable to the number of syllables per second in spoken English and so is suitable for  listeners (see Wood  \cite{Wood:1973} for a detailed view of the information aspects of tempo). However, each transient event's duration will be approximately 22 ms appearing as a band-limited impulse with a cut-off frequency at around 50 Hz, which is below the most sensitive range of the human auditory system. If the playback rate were raised by, say, a factor of 10--20,  the individual impulses would be shifted to a more perceptible frequency range, but at the cost of an indiscernible temporal structure of the impulse series. Thus, pure audification is a trade-off between the macroscopic time scale and the frequency range of the relevant information.

\section{Direct Segmented Sonification (\DSon)}
Following Rohrhuber's approach \cite{Rohrhuber:2010} the \DSon\ process is regarded as a mapping operation between data domain and sound domain. Because the sonification time domain will often be different from the data time domain (e.g., choosing to listen to a 100 s data set over a period of only 10 s) Rohrhuber proposed superscribing sonification domain variables with a ring to distinguish them from data domain variables, thereby enabling the construction of unambiguous mixed domain expressions. In this scheme the sonification operator $\s{\mathcal{S}}$ maps from the given data space $\mathbb{D}$ to the sound signal space $\s{\mathbb{Y}}$: 
\begin{equation}
\s{\mathcal{S}}: \mathbb{D} \mapsto\s{\mathbb{Y}}
\label{eq_sonifoperator}
\end{equation} 
The relation is more explicit at the level of the variables \cite{Vogt:2012}:
\begin{equation}
\s{\mathcal{S}}: x(\td) \mapsto\s{y}\left({\s{t}}, x(\td); \s{p}\right)
\label{eq_sonifoperatorvars}
\end{equation}
The sonification signal $\s{y}$ depends on ${\sont}$ (sonification time), because sound is a temporal phenomenon, on the data $x(\td)$ to be sonified which itself is assumed to depend on a data domain time $\td$, and the parameters $\s{p}$ of the sonification method which determine how the sonification sounds.

\subsection{Sonification Variables}
\begin{table}[]
\caption{Direct Segmented Sonification Variables, Functions, and Operators}
\begin{center}
\begin{tabular}{rll}
\toprule
\textbf{Variable}&\textbf{Description}&\textbf{Value range}\\
\midrule
\textbf{Temporal}& &\\[1mm]
$\comp$ &  time compression factor&sonification duration\\&& $\sonT = T/\comp$.\\
$ \dilate$ &  dilation factor&$\dilate\geqslant 0$\\[2mm]
\textbf{Pitch}& &\\[1mm]
$\s f_\rf$ & reference frequency& \\
 $\scalea,\scaleb$ &  pitch scaling factors& $\langle\scalea,\scaleb\rangle\geqslant 0$\\[2mm]
\textbf{Loudness} & &\\[1mm]
 $\distort$ & power law distortion factor& $\distort\geqslant 1$\\
 $\thresh$ &  amplitude threshold& $\thresh\geqslant 0$\\
 $\gain(..)$ &  gain function & e.g. mean, rms, \ldots\\
 $\decay$ &  decay parameter &\\[2mm]
\textbf{Timbral} & &\\[1mm]
 $\timbre\langle..\rangle$ &  operator for timbral control &e.g., wave shaping,\\&& additive synthesis\\
\bottomrule
\end{tabular}
\end{center}
\label{tab:vars}
\end{table}%
The proposed sonification method uses the variables shown in Table \ref{tab:vars}. The sonification parameter set is then given as $\s{P} = \{\comp, \dilate, \s f_\rf, \scalea, \scaleb, \distort, \thresh, \gain(..), \decay, \timbre\langle..\rangle\}$  with any appropriate subset $\s{p} \subseteq\s{P}$ being used in the models described below. 
The meanings of these variables are given in the sections that follow.
To distinguish sonification time from data domain time, sonification time variables are given as $\sont, \sont_i, \sonT$ and data domain time variables as $\td, \td_i, T $. 

\subsection{General Framework of \DSon}
\label{sec:framework}
\DSon\ relies on the assumption that a one-dimensional time-varying data stream, $x(\td)$, can be subdivided into short non-overlapping segments of generally different length where each segment contains a consistent portion of application-dependent significant information. Thus, identification of the appropriate cutting points is crucial. For example, if one is interested in the short-term fluctuation of a stock price, the crossing points of the actual stock price with a moving average might be a good choice. 
We consider a data stream as a time varying signal $x(\td)$ expressed as a sequence of sampled values $x(n)$ at a sampling rate $f_{\text{s}}$. The duration of the data stream is $T$ seconds, hence the sequence $x(n)$ consists of $N=T\times f_{\text{s}}$ samples. Assuming that the \DSon\ of the data should last for approximately $\sonT$ seconds (the reason for the duration being approximate is explained below), a time compression factor $\comp$  is defined by $\comp =T/\sonT$. 
 
As a first step, the cutting points $\td_i$ (the borders between segments $x_i(\td)$) have to be determined depending on the application and the specific properties of the data. As a simple example, consider a broadband AC signal. In this case the zero crossing points are a reasonable choice. If the signal contains DC or strong low-frequency components (as is the case with stock prices and the data used in \S\ref{sec:FRED}) some preprocessing might be necessary. For instance,  the trend signal $x_{\trend}(\td)$ calculated by a moving average filter  can be subtracted from the original data yielding a signal $x_{\AC}(\td) =x(\td)-x_{\trend}(\td)$ which exhibits numerous zero crossings.

Assuming the first cutting point is at $\td_0=0$ and the last one is at $\td_M=T$, a sequence of $M$ segments $x_i(\td)$ (or $x_{i,\AC}(\td)$ if the low frequency mean or DC component has been removed through preprocessing) is obtained where:
\begin{eqnarray}
x_i(\td) = \left\{\begin{array}{ll}
x\left(\td+\td_{i-1}\right) & 0 \leqslant \td \leqslant (\td_{i}-\td_{i-1})\\
0 & \text{else}
\end{array}\right.
\end{eqnarray}
Thus, the actual duration of each segment is given by  $T_i=\td_i-\td_{i-1}$.
Each data segment $x_i(\td)$ is to be sonified as an individual sonic event $\s{y}_{i}(\sont)$ depending on the parameters $\s{p}$ of the sonification method at hand and is superimposed to form the final sonification $\s{y}(\sont)$. For the sake of simplicity, we skip the explicit dependence of the sonic event $\s{y}_{i}(\sont)$ on the data segment $x_i(\td)$ and the sonification parameters $\s{p}$ in the following:
\begin{eqnarray}
	\s{y}(\sont) =  \sum_{i=1} ^M   {\s{y}_{i}\left(\sont-\s t_{i-1}\right)} & \text{where} &\s t_{i-1}=\dfrac{t_{i-1}}{\comp}
\end{eqnarray}
Note that the individual sonic events $\s{y}_{i}$ might be longer or shorter than the duration of the respective data segment $T_i=\td_i-\td_{i-1}$ depending on the specific sonification method and parameters. Therefore, the actual length $\sonT$ of $\s{y}$ is only approximately equal to the data duration divided by the compression factor: $\sonT \approx T/\comp$. 

The  \DSon\ approach conserves the overall temporal structure of the data as long as the cutting points are chosen appropriately, that is, they are meaningful within the context of the data domain. Since the sonification length of the individual segments is not predetermined by this very general formulation, the resulting auditory display can be adjusted either to focus on the rhythmical structure of the segments' temporal distribution (such as by choosing very short and transient sonic events for each segment and thereby presenting, essentially, a sequence of clicks) or to zoom into the specific data evolution of each segment (e.g., by choosing long sonic events with time-varying properties according to the segment's data values). Note that the latter approach yields a temporal overlap of sonic events of adjacent segments and hence might confound the auditory gestalts originating from the individual segments.
 In any case, the appropriate choice of the sonification method for the individual segments is crucial for the quality of the \DSon. In the following section, a simple method for  segment sonification which is derived from auditory graphing is presented.

\section{Modified Auditory Graphs for Sonifying Individual Segments}
 \label{sec:modAG}
Auditory graphs have been a part of the standard repertoire of auditory display research since its beginning. At its simplest, an auditory graph represents the ordinate value of a data series as the time-varying frequency of a sinusoid with (usually) constant amplitude \cite{Flowers:2005}. An obvious benefit is the  straightforward analogy to visual graphs, which makes them readily understandable, at least for sighted users. 
Flowers \cite{Flowers:2005} recommended using distinct timbres in order to minimize stream confusions and unwanted perceptual grouping. Since auditory graphs usually encode data values as pitch or (fundamental) frequency, harmonic complexes with a small number (6--8) of partials and amplitudes in inverse proportion to partial order are  recommended  instead of pure sinusoids because of the improved pitch salience they are able to produce. Nevertheless, the resulting timbre should be time-invariant to guide the listener's attention to the pitch contour and not obscure the data representation by arbitrary timbral fluctuations. More complex timbres run the risk of evoking categorical associations with real-world sound that might change at more or less arbitrary data values and therefore confound the intended perceptual continuum of the frequency or pitch range representing the  important aspects of the data. If several auditory graphs are to be presented simultaneously spectral overlap between adjacent graphs should be avoided, therefore pure sinusoids might be the better choice in this instance.
 
In order to achieve the intended directness of the final sonification, not only must the overall temporal relationship of the segmentation pattern be preserved (as is ensured by the general framework in \S\ref{sec:framework}), but the sonic events resulting from the individual segments must also display the segments' data evolution as directly as possible. Therefore, a modified auditory graph is proposed as the specific method of segment sonification in \DSon\ with each segment being treated as an individual graph. We assume segments are derived from zero crossing points (either due to the inherent AC characteristics of the data or after removing the signal average) and exploit the property that each segment starts and ends with data values of negligible magnitude. To accentuate strong deviations from a chosen baseline (such as the average), amplitude modulation derived from the segment's data complements the time-varying pitch progression of the basic auditory graph. Thus, the general form of the sonification signal $\s{y}_{i}(\sont)$ is given by
\begin{equation}
\s{y}{_i}(\sont) = a_i({\sont}) \sin\left(2\pi\int_0^{\sont} f_{\rf}\cdot 2^{b_i(\sont')}\cdot \dif\sont'\right)
\end{equation}
where $a_i({\sont})$ is the amplitude modulator, $f_{\rf}$ is the base frequency for the pitch range of the sonification, and  $b_i({\sont})$ is a pitch modulator.%
\footnote{In order to allow specific control of timbre, an additional timbre operator $\timbre$ which acts on the sine function has to be considered in the model:
$$ \s{y_i}(\sont) = a_i({\sont}) \timbre \left\langle \sin\left(2\pi\int_0^{\sont} \s f_{\rf}\cdot 2^{b_i(\sont')}\cdot \dif\sont'\right)\right\rangle$$
$\timbre$ might be implemented as, for instance, waveshaping utilizing Chebyshev polynomials or any kind of additive synthesis. The operator properties itself will depend on the data to be sonified, i.e. $\timbre\langle\sin(..); x_i\rangle$. However, in the case of the modified auditory graph the resulting sonic events consist only of amplitude and pitch modulated sinusoids, hence $\timbre$ can be regarded as the identity function, $\timbre\langle\sin(..); x_i\rangle=\sin(..)$, and will be omitted in the following for the sake of simplicity.} %
 To include the (previously removed) short-term average value as an overall pitch trend, we explicitly take into account both the mean-free segment $x_{i,\AC}(\td)$ and the trend signal at the segment's starting point $x_{\trend}\left(\td_{i-1}\right)$ for pitch modulation.
\begin{IEEEeqnarray}{lll}\label{eq:dilate}
\nonumber\s{y}_{i}(\sont) &\ =\ & \abs{x_i\left(\dilate\cdot {\sont}\right)} \times\\
&&\sin \left(2\pi\int_0 ^{\sont} {f_{\rf}\cdot 2^{\left(x_{\trend}\left(\td_{i-1}\right)+x_{i,\AC}\left(\dilate\cdot \sont'\right)\right)} \dif\sont'}\right)
\end{IEEEeqnarray}

In (\ref{eq:dilate}), the magnitude of the segment's data values is used as amplitude modulation and the dilation parameter $\dilate$ determines the length of the sonic event $\s{T}_i$ in relation to the duration of the data segment $T_i$.  If $\dilate = \comp$, adjacent sonic events do not overlap since  $\s{T}_i =  {T_i}/ \comp $, whereas $\dilate  \leqslant\comp$ results in overlapping events. 
    
 Of course, both pitch and amplitude modulation can be parameterized in various ways. For example, if mainly peak or strong deviations from the mean are to be displayed, a power law distortion $\distort$ can be applied to the amplitude modulator $a$:
 \begin{eqnarray}
    \label {AM_p} 
    a_i(\sont) =  \abs{x_i\left(\dilate\cdot \sont\right)}^{\distort}; &  \distort \geqslant 1
\end{eqnarray}
If only deviations exceeding a threshold $\thresh$ around the mean are to be sonified, then a magnitude offset followed by half-wave rectification might be included in the amplitude modulator:
\begin{eqnarray}
\label {AM_th}a_i(\sont)&=&G\left(\abs{x_i\left(\dilate\cdot\sont\right)},\thresh\right)\\
G(x,\thresh)&=&\left\{\begin{array}{ll}
x - \thresh & x \geqslant \thresh\\
0 & \text{else}
\end{array}\right.
\end{eqnarray}
On the other hand, the relative importance of the trend signal $x_{\trend}$ and the actual data progression of the segment can be 
 adjusted via non-negative parameters $\scalea$ and $\scaleb$.
\begin{equation}
    b_i(\sont')=  \left(\scalea \cdot x_{\trend}\left(\td_{i-1}\right)+\scaleb\cdot x_{i,\AC}\left(\dilate\cdot \sont'\right)\right)
\end{equation}
	
 If the stream of segments with positive deviation from the trend should be discriminated from the stream of negative segments, two different reference frequencies $f_{\rf}^{\up}$ and $f_{\rf}^{\down}$  could be used.  From the above, the general parameterized form of \DSon\ is
\begin{IEEEeqnarray}{lll}
   \nonumber\s{y}_{i}(\sont) &\ =\ & \abs{x_i\left(\dilate\cdot \sont\right)}^{\distort}\times  \\
    \nonumber&&\sin \left(2\pi\int_0 ^{\sont} {f_{\rf}^{\up/\down}\cdot 2^{\left(\scalea \cdot x_{\trend}\left(\td_{i-1}\right)+\scaleb\cdot x_{i,\AC}\left(\dilate\cdot \sont'\right)\right)} \dif\sont'}\right).\\
\end{IEEEeqnarray}

\subsection{Modulation of Segment Duration}
To relate the duration of sonification segments to some property of the data we can use $\dilate$ not as a constant, but as a function of the segment's data, $\dilate_i$. For instance, if highly peaked segments should be displayed as longer sonic events to display the data distribution in more detail, a monotonically decreasing, concave function of the segment's mean (or other property such as rms, power) or area  (or energy) is more suitable for $\dilate_i$.

\subsection{Decaying Envelope as Amplitude Modulator}
In order to emphasize the rhythmical patterns induced by the temporal distribution of the cutting points, a sharp attack of the individual sonic events is needed. This can be achieved by replacing the amplitude modulator $\abs{x_i(\dilate\cdot \sont)}$ or the variants in (\ref{AM_p}) and (\ref{AM_th}) by an appropriate envelope, for example,  $\s g_i \cdot e^{-\sont/\decay}$ or $ \s g_i \cdot \sont \cdot e^{-\sont/\decay}$, where $\decay$ is the envelope's decay parameter and  the gain factor $\s g_i$ is determined by a specific function of the segment's data values, $\s g_i=\s g(x_i)$, e.g., the mean, rms, area, power, or energy of the segment. 

\section{Applying \DSon\ to Biomechanical Data}
\label{sec:FRED}
We applied the above \DSon\ method to biomechanical signal data taken from the Functional Readaptive Exercise Device (FRED), an exercise machine designed for use in physiotherapy to help patients with low back pain \cite{Winnard:2017}. The current version of FRED is a modified cross-trainer but which offers minimal resistance (Fig. \ref{fig:FREDdevice}).  
 This creates a situation in which the user has an unstable base of support: when the front foot comes to the forward-most position in its elliptical path gravity then pulls the foot downward requiring the user to apply compensatory balancing force with the rear foot to control the descent. The goal is to operate the machine with an upright posture in a smooth, controlled manner with as little variability in movement speed as possible \cite{Winnard:2017}.

A rotary encoder in the drive wheel generates a pulse stream which represents the instantaneous angular velocity of the wheel at each sampling point. This pulse stream is sampled at 4 kHz into LabChart \cite{ADInstruments}. The data is converted to  frequency values (i.e., revolutions per second) for ease of display for the user (a patient). The resultant data stream is then smoothed using a triangular Bartlett filter to remove the steps in the data. The smoothed stream is presented to the user  via LabChart (with a zoom level of 50:1) as a means of feedback to help them control their performance (Fig. \ref{fig:live}). 

\begin{figure}[ht]
\begin{center}
\includegraphics[width=\linewidth]{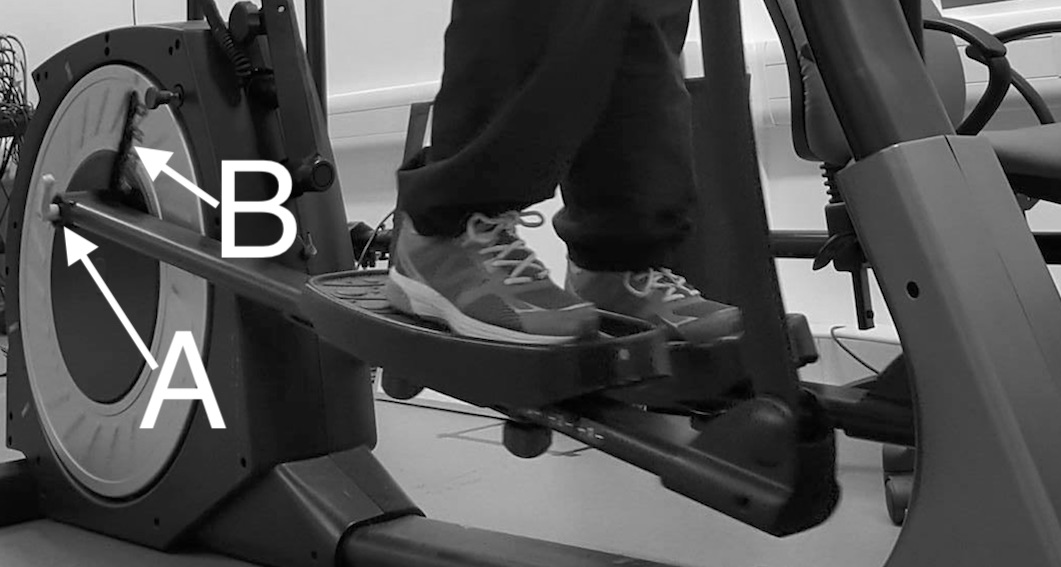}
\caption{The height (difficulty) of FRED's walking path is increased by moving the rear end of the stride rail (A) through the slot (B) towards the edge of the wheel. FRED has five such settings and is shown here in its default (lowest) configuration.}
\label{fig:FREDdevice}
\end{center}
\end{figure}
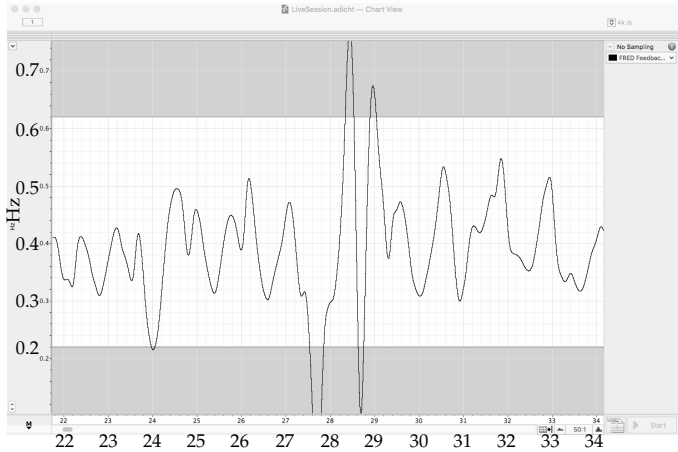
\begin{figure}[ht]
\begin{center}
\labellist
\scriptsize
\pinlabel 22 at 22 -2
\pinlabel 23 at 38.5 -2
\pinlabel 24 at 55 -2
\pinlabel 25 at 71.5 -2
\pinlabel 26 at 88 -2
\pinlabel 27 at 105 -2
\pinlabel 28 at 122 -2
\pinlabel 29 at 138.5 -2
\pinlabel 30 at 155.5 -2
\pinlabel 31 at 172 -2
\pinlabel 32 at 188.5 -2
\pinlabel 33 at 205 -2
\pinlabel 34 at 221.5 -2
\pinlabel \rotatebox{90}{Hz} at 2.5 85
\scriptsize
\pinlabel 0.2 at 8 33
\pinlabel 0.3 at 8 51
\pinlabel 0.4 at 8 72.5
\pinlabel 0.5 at 8 93.5
\pinlabel 0.6 at 8 115.5
\pinlabel 0.7 at 8 138
\endlabellist
\includegraphics{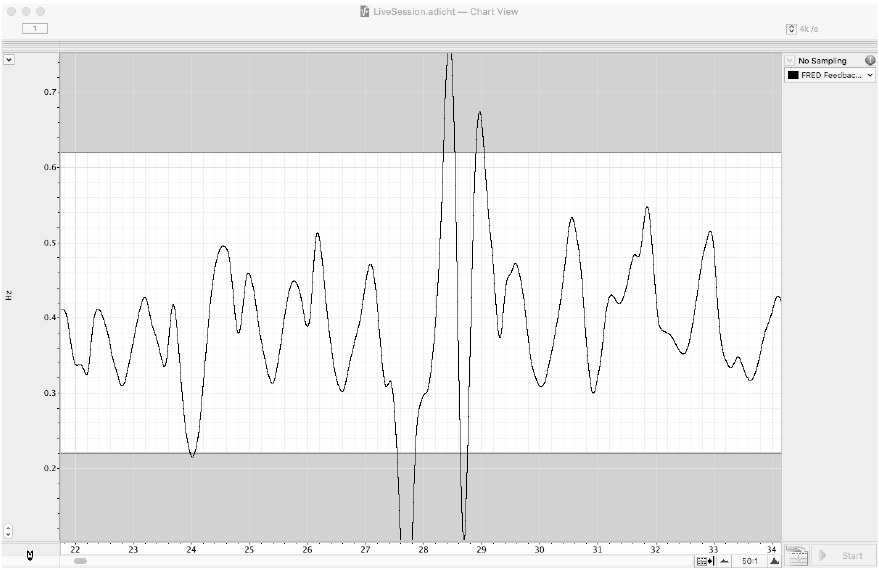}
\caption{Screen capture of the live scrolling window view in LabChart as seen by a FRED user during an exercise session (axis labels have been superimposed here to aid the reader).}
\label{fig:live}
\end{center}
\end{figure}
It has been determined that with the machine in its default configuration (Fig. \ref{fig:FREDdevice}), 
operating it within a frequency range of 0.2 Hz $\leqslant f \leqslant$ 0.6 Hz results in therapeutic benefit leading to recruitment of the key spinal and abdominal muscles \emph{lumbar multifidus} (LM) and \emph{transversus abdominis} (TrA), and with the biomechanical optimum for maximum benefit being achieved at $f=0.4$ Hz \cite{Winnard:2017}. At this optimal frequency a complete rotation of the footplates takes 2.5 s, thus requiring a slow and steady pace. 

The white area in Fig. \ref{fig:live} shows the user when they are performing inside the required range with the shaded areas denoting frequencies above and below the required range. Fig. \ref{fig:live} shows the user is maintaining a good pace until 27.2 s at which point they slow down dramatically, coming to a brief halt (27.65 s) followed by a sharp corrective acceleration which takes the frequency up to 0.8 Hz followed by a compensatory attempt to slow down, followed by another sharp acceleration, with normal performance being re-attained at around 29.2 s. 
\begin{figure}[ht]
\begin{center}
\includegraphics{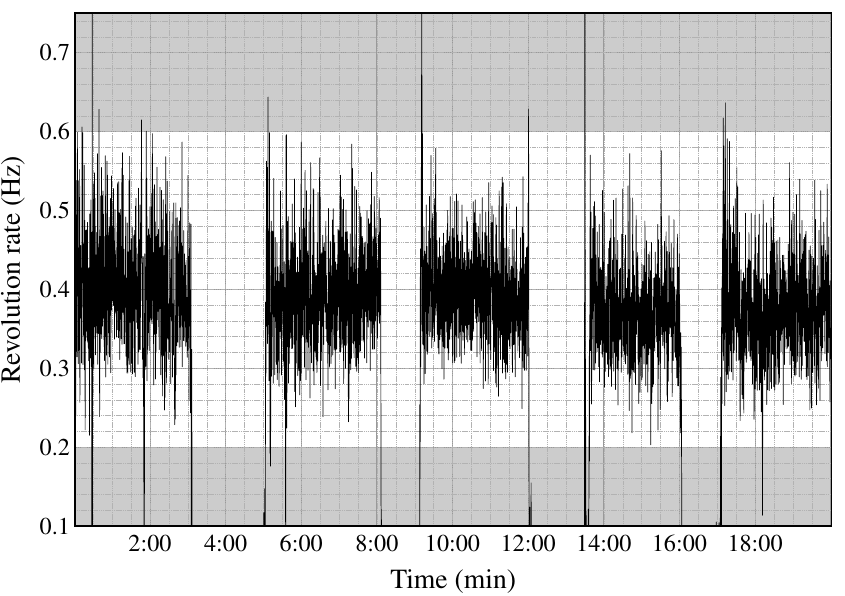}
\caption{A full FRED exercise session comprising one three-minute warm-up block and four three-minute exercise blocks with rest periods.}
\label{fig:session}
\end{center}
\end{figure}

A typical  session comprises several three-minute blocks of exercise separated by rest periods of between 30--60 s. A full session's worth of data is shown in Fig. \ref{fig:session}. A quick glance at the data plotted at a zoom level of 2000:1 (Fig. \ref{fig:session}) is helpful to the physiotherapist for getting an overall impression of the user's performance. During rest periods the therapist uses this zoomed-out view to look for signs of fatigue (such as a rising trend line in the frequency) which may require extension of the rest period. The plots are useful to the therapist during an exercise session but post-hoc review of many session data files quickly becomes tiring. Repeated zooming in and out is needed to locate regions of possible interest and to spot specific instances of particular performance behaviours.

\subsection{Features of Interest}
\label{sec:features}
\begin{figure}[ht]%
       \subfloat[Double ``flat top'']{%
     \label{fig:flattop}%
     \includegraphics{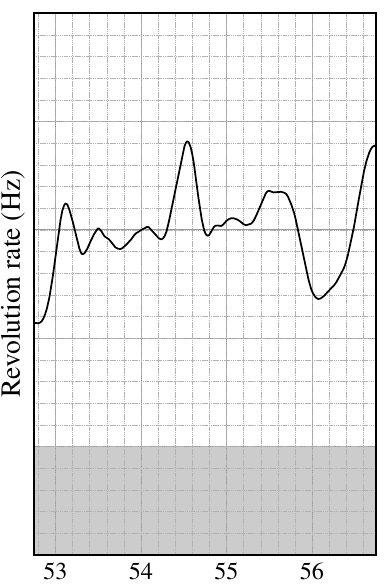}}%
       \subfloat[Controlled recovery]{%
     \label{fig:recovery}%
           \includegraphics{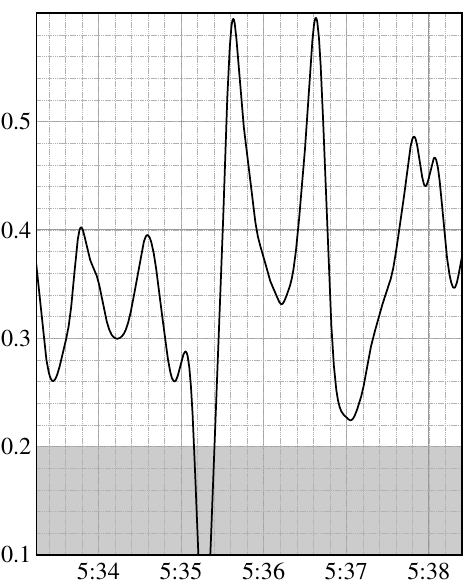}}
       \caption{Strong and weak performance. In (a) the user demonstrates two periods of very small deviation in velocity. In (b) the user's velocity drops below target but is very quickly recovered back into the target range.}%
       \label{fig:FlatTopRecovery}%
\end{figure}

During a post-hoc review of performance, the physiotherapist is interested in identifying a number of discrete features in the data sets. The main performance goal is to maintain a walking pace of 0.2 Hz $\leqslant f\leqslant$ 0.6 Hz. While the patient needs to be aware of excursions outside this range during exercise, for the therapist all excursions above 0.6 Hz and long excursions below 0.2 Hz are of interest. If the frequency exceeds 0.6 Hz (Fig. \ref{fig:live}) this indicates a loss of control --- the machine is running away with the user. However, because it takes a great deal of muscle control to operate the machine slowly, if the frequency momentarily drops below 0.2 Hz and then goes back in range this is of less interest to the therapist as it is still evidence of control --- it is a controlled recovery (Fig. \ref{fig:recovery}). But if it drops below 0.2 Hz for an extended period of time (typically half-a-second or more) then this also indicates a lack of control as motion is coming to a stop.

The target range of 0.2 Hz $\leqslant f \leqslant$ 0.6 Hz means that users can demonstrate variability in their average speed while still maintaining acceptable performance. Therefore, for each user, the physiotherapist will additionally determine a  maximum deviation from the individual mean as a target range based upon their assessment of the user's current ability and any physical characteristics that might impact upon how well they are able to use FRED. For example, a beginner with reasonable control might be expected to achieve a standard target deviation of 0.15 Hz while someone who is able to keep within the range 0.35 Hz $\leqslant f\leqslant$ 0.45 Hz would have a target deviation of  0.05 Hz. Once the therapist has determined a user's target deviation it is interesting to know at what points they are failing to maintain it.

If someone were able to operate the machine perfectly there would be no variation in their speed and the plot would show a flat line. Therefore, the smoother the plot the less the user's pace is varying. When a user starts to master the required walking technique they begin to exhibit what are known as ``flat tops''.  A flat top is a region of activity lasting approximately 0.5 s or more in which the variation in speed is so small that the curve starts to flatten out. Flat tops typically occur during the portion of a walking cycle after the rear foot has come up from the bottom of the elliptical path and before the front foot descends again. Fig. \ref{fig:flattop} shows a double flat top. At around the 53 s mark the small peak indicates where the user's rear foot has ascended from the bottom of the elliptical path. This is followed by a period of relatively flat speed variation lasting just under 1 s. At around 54.2 s the front foot descends and then another flat top of  $\approx$0.7 s occurs.

Because these features require zooming in to see clearly it becomes time consuming to zoom-and-scroll through many data files, so  \DSon\  was applied to FRED data sets to see how well these features could be heard. After discussions with physiotherapists from Northumbria University's Aerospace Medicine and Rehabilitation lab in which FRED is being further developed, the features to be represented were:

\begin{enumerate}
\item Any excursions above 0.6 Hz.
\item Long excursions below 0.2 Hz.
\item Periods  outside the user's target deviation range.
\item `Flat tops' lasting $\approx$0.5 s or longer.
\end{enumerate}

The preprocessing stage involved audifying FRED data streams by simply converting each data point to a signed 16-bit integer and storing the result in a PCM-encoded digital audio file. Because the revolution rate does not exceed 2 Hz (which would be very fast walking) the signal spectrum caused by the speed fluctuations occurring during a full revolution is band limited below 15--20Hz. Therefore, to keep the file sizes small the data extracted from LabChart were first downsampled to $f_{s}=100$ Hz prior to audification. 
 Thus, the time series signal, $x(\td)$ in the \DSon\ method was provided by these audio files. The \DSon\ method was implemented in a series of MATLAB (for sonification) and Python (preprocessing) scripts (see the project repository \cite{nuson-DSSon}).

\section {\DSon\ Models for FRED Signals}
In this section we describe three \DSon\ models that were applied to FRED data that emphasize the features of interest identified above to varying degrees resulting in differing auditory saliency. \DSon\ for FRED data is mainly intended to provide an auditory display of users' performance that enables the physiotherapist to conduct a quick analysis during post-hoc review. The \DSon\ parameters might also be individually adjusted by the therapist during the review session in order to concentrate on specific data features. Consequently, it is impractical to evaluate the \DSon\ display through extensive listening tests based on specific task completion performance  and statistical analysis. This kind of evaluation procedure is planned for future work on other application fields. Here, \DSon's properties (benefits and limitations) are demonstrated  by comparing data excerpts containing specific features of interest and the resulting \DSon\ display. 
Audio files, demonstrating the system output, together with the corresponding data sets used to generate them, can be found in the project repository \cite{nuson-DSSon} and are listed in  Table \ref{tab:audio}.
\begin{table}[htbp]
\caption{Example Sound Files}
\begin{center}
\begin{tabular}{rlll}
\toprule
\textbf{\#}&\textbf{Audio file}&\textbf{Description}\\
\midrule
1&\texttt{DSSon\_Basic\_A\_n.wav}& M1,  user A --- novice\\
2 &\texttt{DSSon\_Basic\_A\_e.wav}& M1, user A --- experienced\\
3 &\texttt{DSSon\_Basic\_B.wav}& M1, user B --- novice\\
4 &\texttt{DSSon\_ITR\_A\_e.wav}& M2, user A --- exp.\\
5 &\texttt{DSSon\_ITR\_B.wav}&M2, user B --- novice\\
6 &\texttt{DSSon\_ADV\_A\_n.wav}& M3, user A --- novice\\
7 &\texttt{DSSon\_Adv\_A\_e.wav}&  M3, user A --- exp.\\
8 &\texttt{DSSon\_Adv\_B.wav}&  M3, user B --- novice\\
\cmidrule{1-3}
\multicolumn{3}{L{0.9\linewidth}}{Models: M1 = basic model; M2 = individual target range model; M3 = advanced model}\\
\multicolumn{3}{L{0.9\linewidth}}{ Data files used: user A, novice = DA1; user A, experienced = DA2; user B, novice = DB1}\\
\bottomrule
\end{tabular}
\end{center}
\label{tab:audio}
\end{table}%

\begin{figure}[htbp]
\begin{center}
\includegraphics[width=\linewidth]{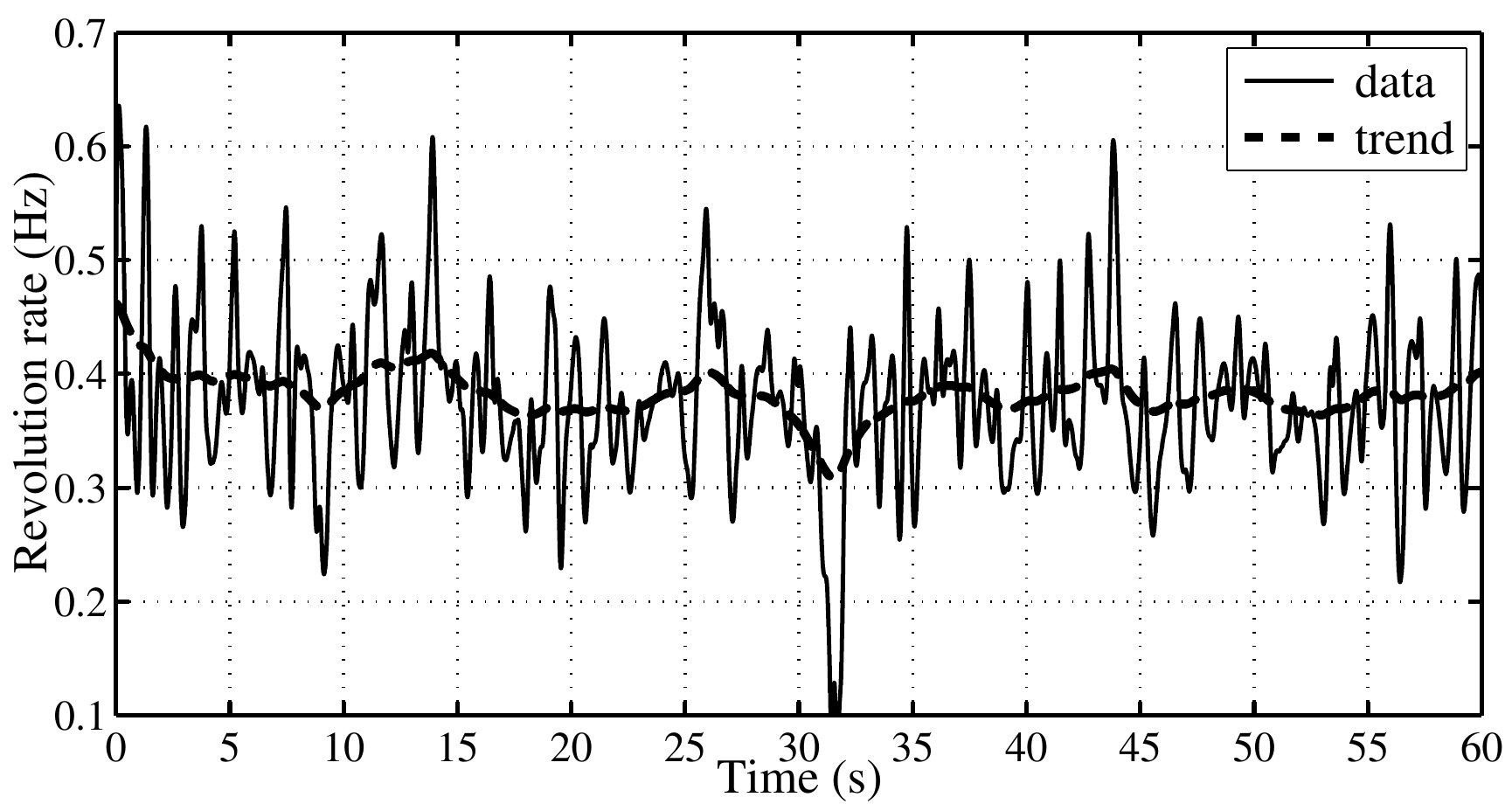}
\caption{Data stream and trend signal  (weighting factor $w=0.2$) of FRED exercise sessions of user A at the beginning of training (audio file 1).}
\label{fig:trend_DA1}
\end{center}
\end{figure}

\begin{figure}[htbp]
\begin{center}
\includegraphics[width=\linewidth]{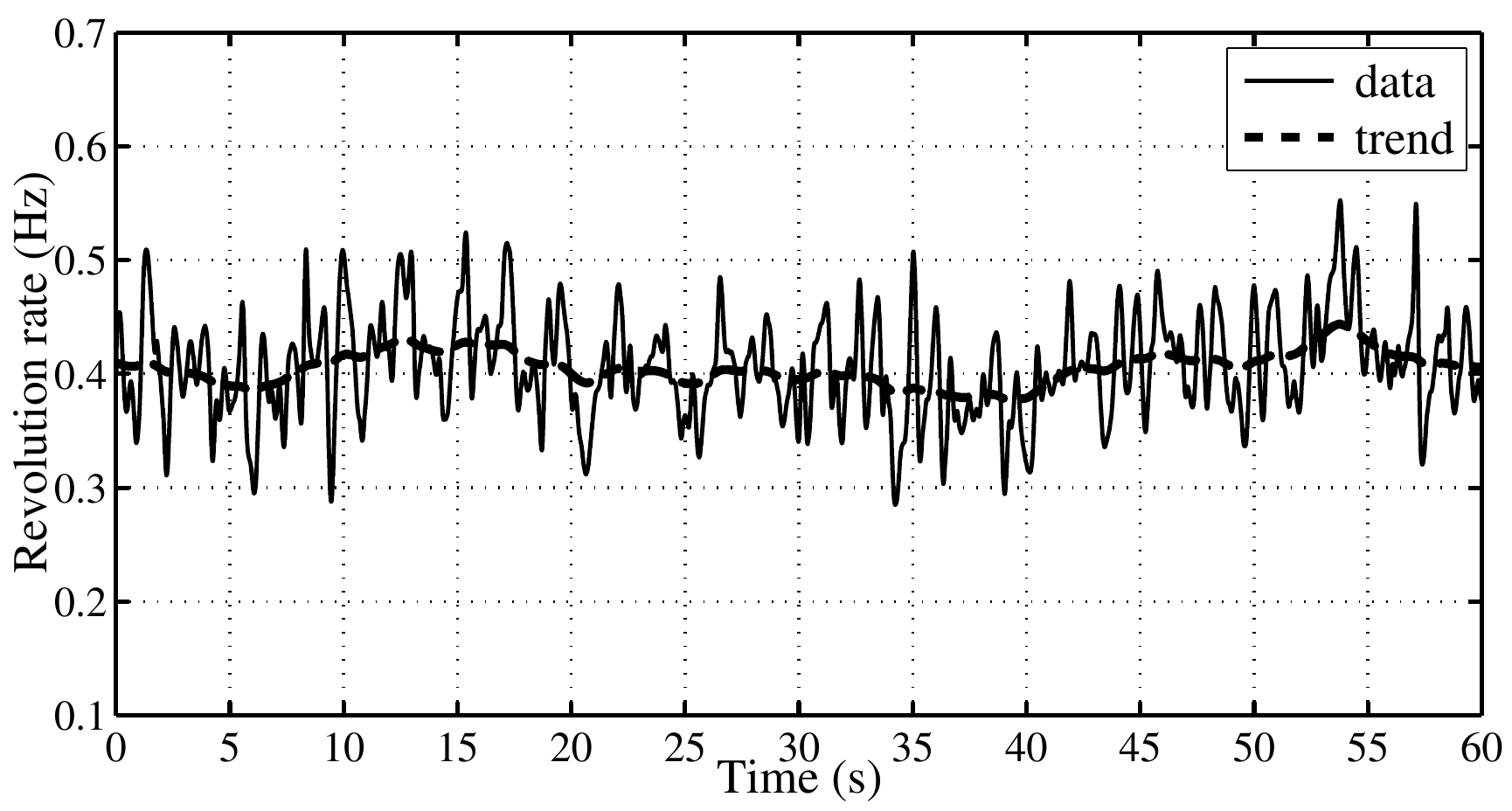}
\caption{Data stream and trend signal  (weighting factor $w=0.2$) of FRED exercise sessions of user A, four months after training (audio file 2).}
\label{fig:trend_DA2}
\end{center}
\end{figure}

The first step in \DSon\ is signal segmentation. For FRED data, the main feature of interest is the deviation of the instantaneous revolution rate from the fixed target value, $x_{\text{target}}=$ 0.4 Hz (the biomechanical optimum from above). Hence, an obvious choice for segmentation is to cut the data stream at its crossing points with this target value, that is, extract segments with positive and negative deviation from  $x_{\text{target}}$. However, as far as a user is able to maintain a steady revolution rate, even slightly deviating from 0.4 Hz, or shows a slowly varying average revolution rate exhibiting only small excursions, he/she shows sufficient muscle control and therefore gains therapeutic benefit. To account for this fact, we did not use the fixed target value of 0.4 Hz to determine the segments' start and end points, but calculated a weighted mean of the target and the moving average of the data stream,  $x_{\text{MA}}(t)$, to obtain the trend signal:
 $$x_{\trend}(t) = w \cdot x_{\text{target}} + (1-w) \cdot x_{\text{MA}}(t).$$
The data stream and the trend signal (weighting factor $w=0.2$) of two exercise sessions of the same user are shown in Figs. \ref{fig:trend_DA1} and \ref{fig:trend_DA2}. The first data stream was recorded in the second week of a six-week training period, and the second was recorded four months after the end of the training period. The data segments are determined utilizing the zero-crossing points of the trend-free signal:  
$x_{i}(t)=x(t-t_{i-1})-x_{\trend}(t-t_{i-1})$ for $ t_{i-1} \leqslant t \leqslant t_i$, and $x_i(t)=0$ otherwise. 

\subsection{\DSon\ Basic Model}
The \DSon\ basic model uses a time compression factor $\comp=t/\s t=5$ and a dilation parameter $\dilate =T_i/\s T_i=5$. This moderate compression factor allows for a rather fast post-hoc review of the data. The sonic events resulting from adjacent positive and negative excursions are displayed at a rate of approximately 8 events per second, that is, a mean revolution rate of 0.4 Hz times (typically) 4 segments per revolution (2 positive and 2 negative excursions) times compression factor $\comp=5$. This rhythmical pattern can be easily perceived in detail because it lies quite within the typical range of musical gestures and the individual events do not overlap due to the dilation parameter chosen ($\dilate =\comp$). In order to better facilitate the discrimination between positive and negative excursions, different reference frequencies for the pitch modulator are employed, specifically $\s f_{\rf}^{\up}=400 \text{~Hz}$ and $\s f_{\rf}^{\down}=300 \text{~Hz}$. To monitor both the individual excursions and the overall trend, both pitch scaling factors are applied $\scalea=\scaleb=2$. Amplitude modulation derived from the instantaneous magnitude of the segment's data values is used, that is, the power law distortion factor $\distort$ equals $1$. 
The final model including the parameter values reads:
\begin{IEEEeqnarray}{lll}
	 \nonumber\s{y}_{i}(\sont)&\ =\ &\abs{x_i\left(5 \sont\right)}\times\\
	 \nonumber&&\sin \left(2\pi\int_0 ^{\sont} 
	 {^{\up:400\text{ Hz}}_{\down:300\text{ Hz}}
	 \cdot 2^{\left(2 \cdot x_{\trend}\left(\td_{i-1}\right)+2\cdot x_{i,\AC}\left(5\sont'\right)\right)} \dif\sont'}\right)\\
\end{IEEEeqnarray} 
\begin{equation}
	\s{y}(\sont) =  \sum_{i=1} ^M   {\s{y}_{i}\left(\sont-{ \dfrac {\td_{i-1}}{5}}\right)}
\end{equation}

\begin{figure}[ht]
\begin{center}
\includegraphics[width=\linewidth]{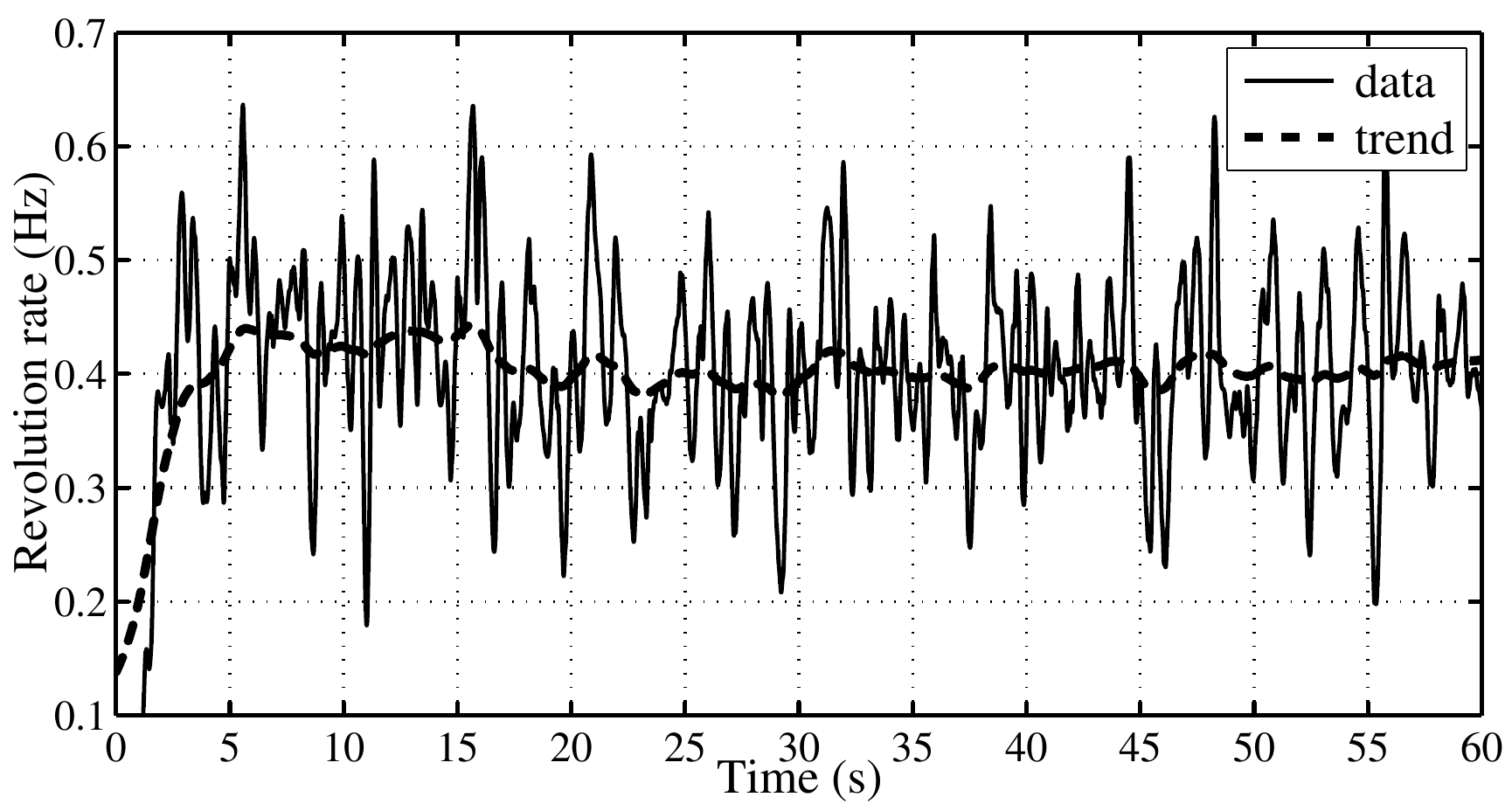}
\caption{Data stream and trend signal  (weighting factor $w=0.2$) of FRED exercise session for a poor performer (user B).}
\label{fig:trend_DB1}
\end{center}
\end{figure}

\begin{figure}[ht]
\begin{center}
\includegraphics[width=\linewidth]{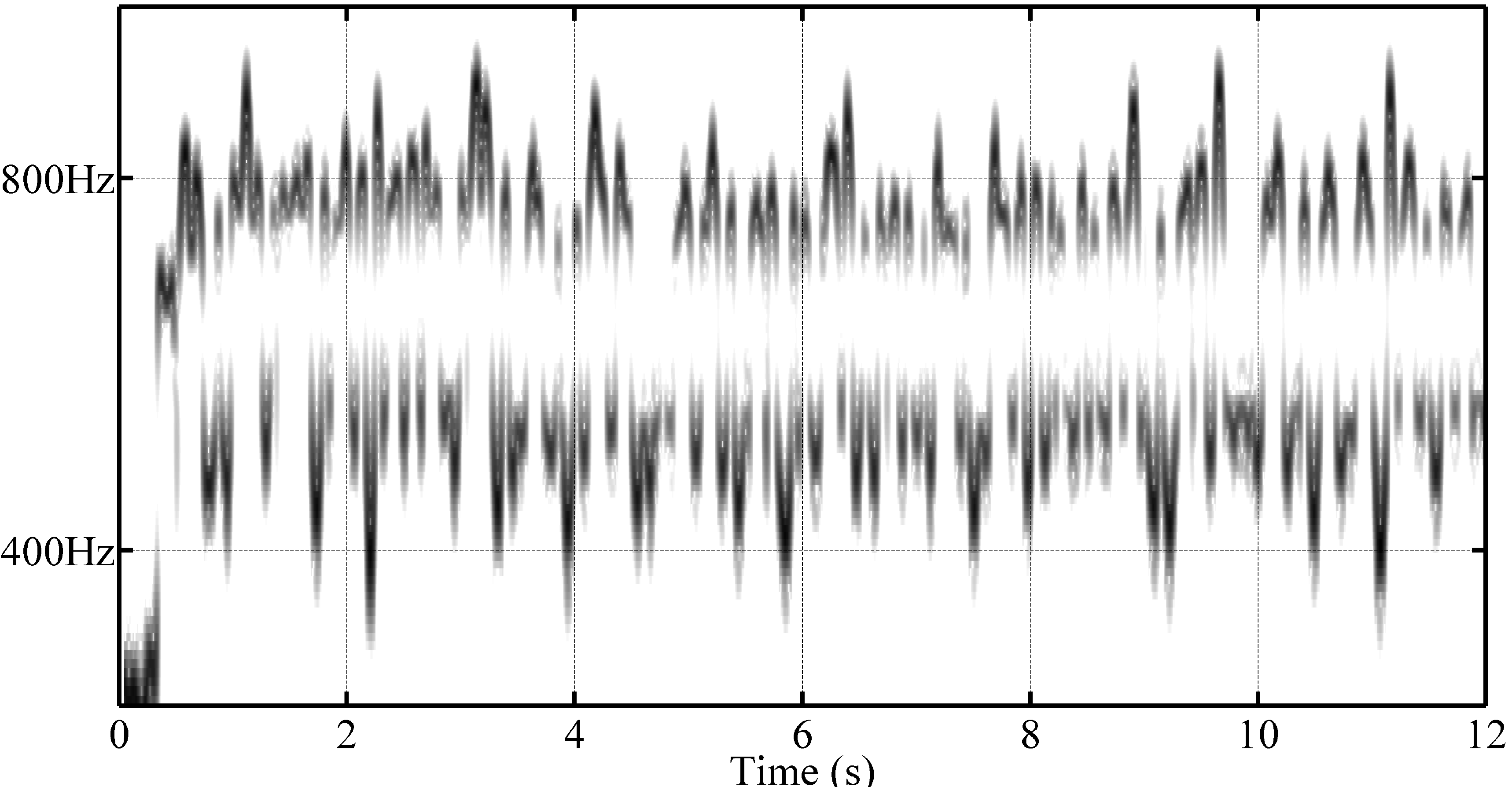}
\caption{\DSon\ basic model spectrogram of User B (Fig. \ref{fig:trend_DB1}, audio file 3).}
\label{fig:spectrogram_DB1}
\end{center}
\end{figure}

The model was applied to three FRED data signals, two from user A (audio files 1, 2) and one from user B (audio file 3). Figs. 
\ref{fig:trend_DB1} and \ref{fig:spectrogram_DB1} show the data and trend as well as the spectrogram of the basic \DSon\ model for a rather poor performance (user B, audio file 3). The user is obviously not able to maintain a stable mean speed at the beginning of the exercise session nor to stay within the range of 0.2 Hz -- 0.6 Hz. Large positive excursions are clearly visible at 6 and 15 s in Fig. \ref{fig:trend_DB1} and result in strong high frequency events at 1 and 3 s (Fig. \ref{fig:spectrogram_DB1}). Sudden slow instants at 11, 45, and 55 s yield prominent low frequency sounds at 2, 9, and 11 s accordingly (Fig. \ref{fig:spectrogram_DB1}). 
Note that highlighting the trend (of approximately 0.4 Hz) in the sonification (due to $\scalea>0$) results in an upward shift of the pitch register compared to the range of the reference frequencies.%
\footnote{If $\scalea=0$, the trend data are completely suppressed resulting in a lower pitch register. If $\scalea=2$ and the trend equals 0.4Hz, then the instantaneous frequencies are multiplied by $2^{\scalea 0.4}=2^{0.8}=1.75$, resulting in a center frequency of  $f =0.5\times(300+400)\times 1.75 \approx$ 610 Hz instead of 350 Hz.} 
The trend variation results in an overall glissando gliding upward and downward displayed in the spectrogram as the sliding white frequency band framed by the sonic events of positive and negative segments respectively.

\begin{figure}[ht]
\begin{center}
\includegraphics[width=\linewidth]{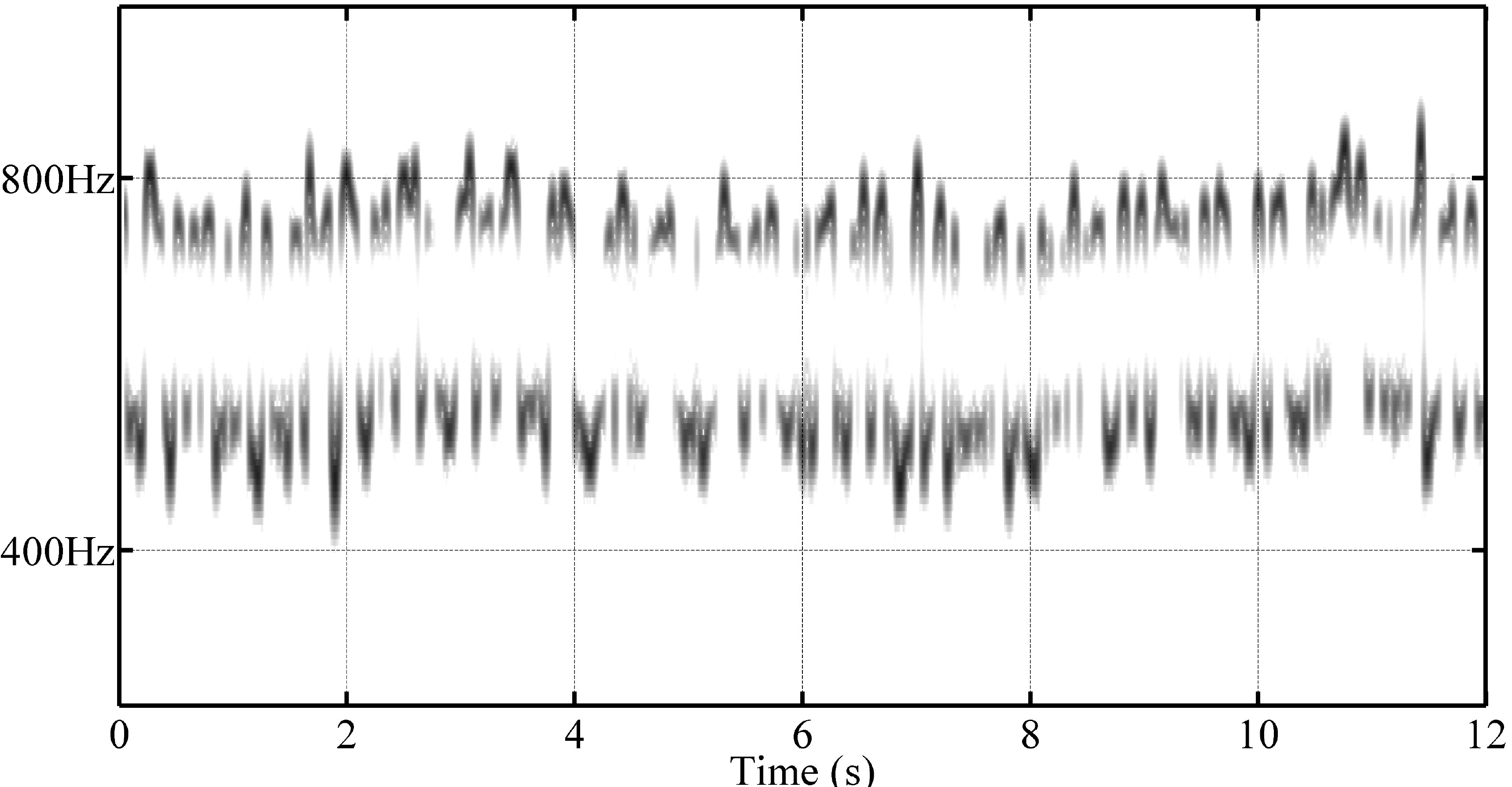}
\caption{Spectrogram of \DSon\ basic model for an experienced user (FRED data and trend are shown in Fig. \ref{fig:trend_DA2}) (audio file 2).}
\label{fig:spectrogram_DA2}
\end{center}
\end{figure}

In comparison, the \DSon\  of the experienced user (Fig. \ref{fig:trend_DA2}, audio file 2) is shown as the spectrogram in Fig. \ref{fig:spectrogram_DA2}. A constant mean rate and regular small deviations resulting in a soft and steady rhythmical pattern dominate this example.

\subsection{\DSon\ Individual Target Range Model}
The time compression factor $\comp=5$ used in the previous examples allows for a quick review of an individual performance. Nevertheless, exploring a collection of FRED sessions consisting of up to five exercise blocks each of 3 minutes duration, would result in a rather time-consuming endeavour and providing sonification with an even larger time compression of  $\comp=10..20$ is preferable. However, the increased playback speed means that the rhythmical patterns of the sonic events and their pitch contours would become indiscernible if the \DSon\ basic model with its previous parameter values were employed. 

Therefore, the \DSon\ individual target range model (ITR) suppresses segments whose maximum excursions stay below the target range set for each user individually by the physiotherapist. This is accomplished by a threshold-based amplitude modulator similar to the  one proposed in (\ref{AM_th}) and setting the threshold parameter $\thresh$ appropriately. Contrary to the amplitude modulator in (\ref{AM_th}) which displays only the segment's data values exceeding the threshold, one might be interested to listen to the entire segment if its value exceeds the target range at some point. Hence, a threshold-based indicator function combined with the segment's instantaneous magnitude is used as the amplitude modulator $a_i(\sont)$:
\begin{eqnarray}
\label{AM_th_2}
a_i(\sont) = \left\{\begin{array}{ll}
\abs{x_i\left(\dilate\cdot \sont\right)} & \max \limits_{\sont}\left(\abs{x_i\left(\dilate\cdot\sont\right)}\right) \geqslant\thresh\\
0 & \text{else}
\end{array}\right.
\end{eqnarray}

\begin{figure}[ht]
\begin{center}
\includegraphics[width=\linewidth]{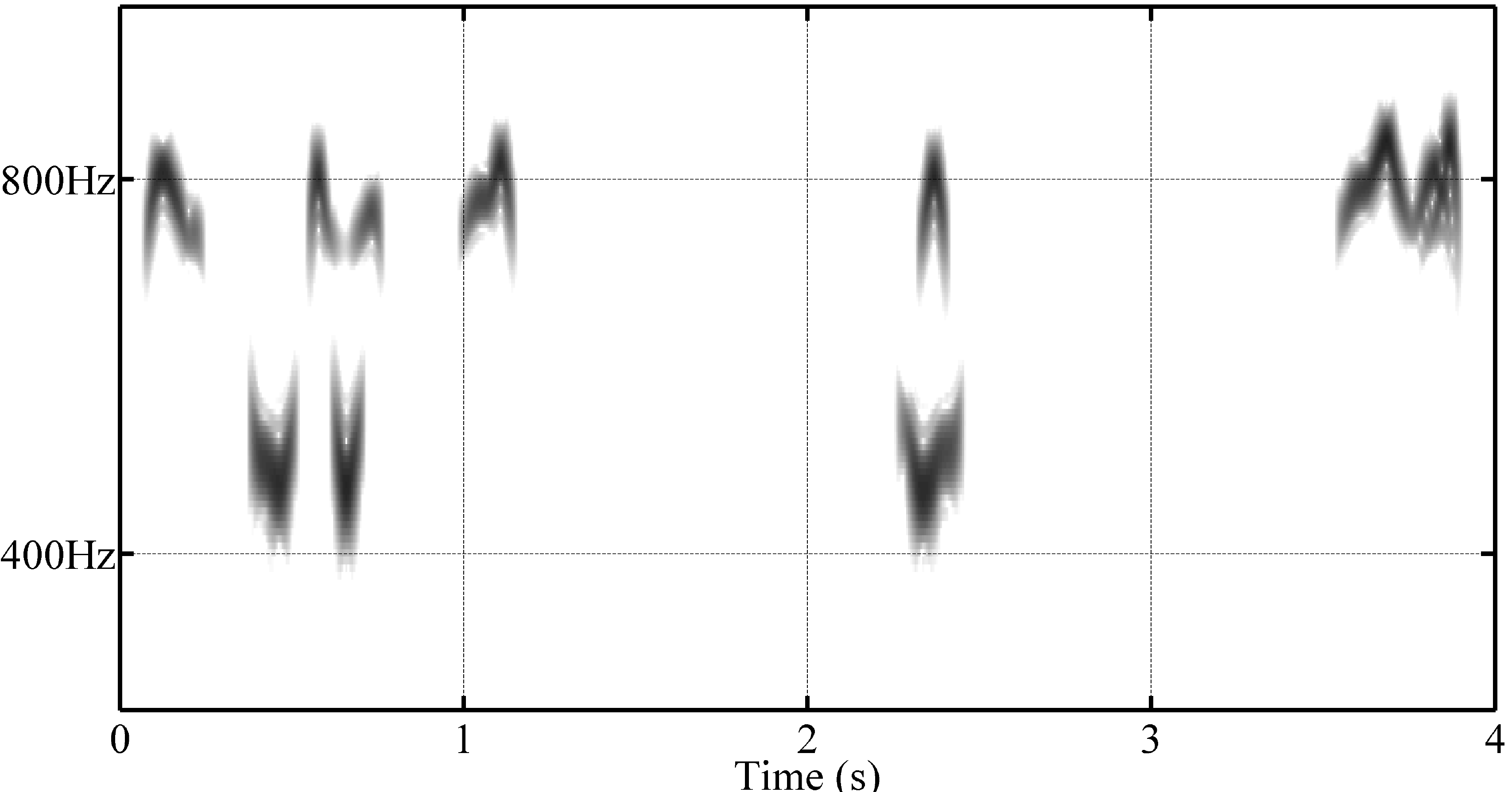}
\caption{\DSon\ ITR model experienced user spectrogram (audio file 4).}
\label{fig:spectrogram_DA2_ITR}
\end{center}
\end{figure}

\begin{figure}[ht]
\begin{center}
\includegraphics[width=\linewidth]{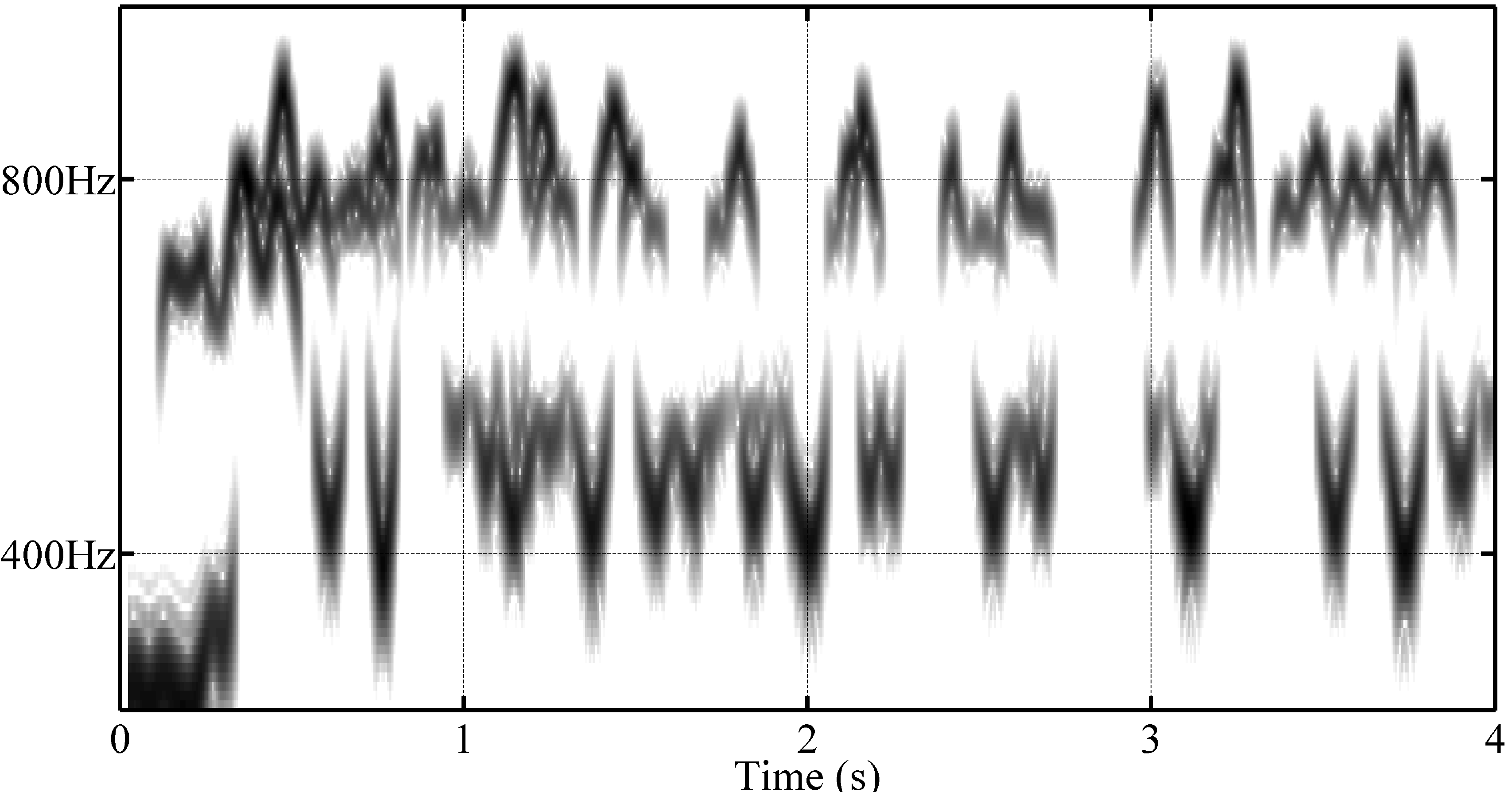}
\caption{DSSon\ ITR model spectrogram for user B (audio file 5).}
\label{fig:spectrogram_DB1_ITR}
\end{center}
\end{figure}

To display the remaining segments in sufficient detail, the dilation parameter $\dilate$ is set as $\dilate< \comp$ yielding potentially  overlapping sonic events. Figs. \ref{fig:spectrogram_DA2_ITR} and \ref{fig:spectrogram_DB1_ITR} show the spectrograms of the new model for the two users. $\comp=15$ results in a sonification duration of 4 seconds for a 1 minute session, $\dilate=5$ yielding a threefold overlap of adjacent sonic events. The threshold parameter $\thresh$ is set to 0.1 Hz for both examples though in practice the therapist would have chosen individual values for the two users according to their level of motor control. All other sonification parameters are set as in the basic model. Note that for the experienced user (Fig. \ref{fig:spectrogram_DA2_ITR}), a sparse auditory display is obtained by the new model (audio file 4)  whereas a dense sonification with almost constantly overlapping sonic events is caused by the poor performance of user B (Fig. \ref{fig:spectrogram_DB1_ITR}, audio file 5). 
\subsection{\DSon\ Advanced Model}
\label{sec:adv}
Both \DSon\ models presented so far are based on a modified auditory graph of adjacent data segments. They are character\-ized by a smooth functional relationship between data values and the auditory display which can be easily perceived by the listener. As every segment is sonified by an amplitude and pitch modulated sinusoid, a coherent auditory gestalt of homogeneous timbre emerges. However, the special features of interest mentioned in subsection \ref{sec:features} are not displayed saliently except for the ITR model which delivers sonic events only for segments exceeding the individual target range, thereby explicitly displaying feature \#3. In order to indicate excursions above 0.6 Hz (feature \#1) and  below 0.2 Hz (feature \#2) prominently, timbre modifications are utilized as an additional sonification parameter. Segments whose maximum excursions cross these limits, are sonified by a fixed harmonic complex (for overshoots above 0.6 Hz) or subharmonic complex (for undershoots below 0.2 Hz) respectively. This is achieved by including the timbre operator $\timbre\langle..\rangle$ in a \DSon\ advanced model (ADV) as:
\begin{IEEEeqnarray}{ll}
\label{timbre}
\nonumber\timbre\langle \sin(\phi_i(\sont))
\rangle = &\left\{\begin{array}{ll}
\s{c}_o\sum_{j=1}^{\sJ}{j^{-\snu}\sin\left(j \cdot \phi_i(\sont)\right)} & \text{o/shoot} \\[3mm]
\s{c}_u\sum_{j=1}^{\sJ}{j^{-\snu}\sin\left(\dfrac{1}{j} \cdot \phi_i(\sont)\right)} & \text{u/shoot}\\
\sin\left(\phi_i(\sont)\right) & \text{else}.\\
\end{array}\right.\\
\end{IEEEeqnarray}
The auxiliary sonification parameters $\sJ$ and $\snu$ specify the number of partials, hence the bandwidth of the sonic event, and the amplitude attenuation associated with increasing partial order. $\s{c}_o$ and $\s{c}_u$ are set so as to align the loudness levels of the overshoot and undershoot segment with the basic one (in this case, $\s{c}_o=0.5$ and $\s{c}_u=0.7$).  Note that by introducing a non-trivial timbre operator, the additional distinct categories of sonic events will result in a sonification where three auditory streams are likely to be perceived and the coherent gestalts of the previous models become dispersed.

To further accentuate segments of long excursions which predominantly occur for undershoots, a data-dependent transformation of the dilation parameter $\dilate$ is incorporated in the ADV model. For data segments whose maximum excursions stay within specified limits (e.g. $0.2 \text{ Hz} \leqslant f \leqslant 0.6 \text{ Hz}$), the dilation parameter is fixed to $\dilate=\dilate_0$, whereas for overshoot and undershoot segments, the dilation parameter becomes a monotonically decreasing function of the segment's data values, $\dilate_i$, and causes stretched sonic events. As a transformation, we specifically propose the hyperbolic 
function of the segment's area, that is, the time integral of segment's magnitude $A_i=\int_{t_{i-1}}^{t_i} \abs{x(t)} \dif t$:  
\begin{eqnarray}
\label{dilate_mod}
\dilate_i = \left\{\begin{array}{ll}
\dfrac {1} {\ssigma} \cdot\dfrac {\sA} { A_i} \cdot {\dilate_0}& A_i \geqslant \sA\\[3mm]
\dilate_0 & \text{else}.\\
\end{array}\right.
\end{eqnarray}
The hyperbolic function translates into a linear dependence of the sonic event's duration $\s{T_i}$ on the segment's area $A_i$, since (\ref{dilate_mod}) and $\dilate_i = T_i / \s{T_i}$ lead to 
\begin{eqnarray}
\s{T_i}= \ssigma \dfrac {A_i} {\sA \cdot \dilate_0 } \cdot T_i & \text{for} &A_i \geqslant \sA.
\end{eqnarray} 
The additional sonification parameters $\sA$ and $\ssigma \geqslant 1$ determine the area threshold and the strength of the dilation transformation respectively. The area threshold should be set to $\sA = {1}/{8 \pi}$ which equals the area of a  sine-formed segment of duration $T= {1}/ {(4 \times 0.4 \text{ Hz}})  $ (the expected duration of an excursion at target revolution rate of 0.4 Hz) and of amplitude 0.2 (magnitude difference between either limit, i.e., 0.2 Hz and 0.6 Hz, and the target rate).
Utilizing this dilation transformation yields dominant stretched sonic events for long overshoot and undershoot segments. However, because the amplitude modulator used up to this point ((\ref{AM_p}) and (\ref{AM_th_2})) delays the loudness peaks of the stretched events, the temporal structure of data segmentation is likely to get obscured. Therefore, an envelope-based amplitude modulation with a rather sharp attack followed by a decay and weighted by the segment's maximum magnitude 
{$x_{i}^\text{max}= \max \limits_{t}\left(\abs{x_i(t)}\right)$ is considered for overshoots and undershoots in the ADV model: 
\begin{eqnarray}
\label{AM_env}
a_i(\sont) =  x_{i}^\text{max} \cdot \dfrac { \dilate_i \sont} {\decay} \cdot e^{- \left(\dfrac {\dilate_i \sont} { \decay}-1\right)}.
 \end{eqnarray} 
The decay parameter $\decay$ is set to $\decay = 0.13 \cdot T_i$ which leads the sonic event to end at an amplitude level of $-40$ dB relative to its maximum. To prevent annoying clicks, a short fade-out portion is further applied at the very end of the envelope. The complete amplitude modulator for the ADV model reads as:
\begin{eqnarray}
\label{AM_ADV}
a_i(\sont) =  \left\{\begin{array}{lll}
 x_{i}^\text{max} \cdot \dfrac { \dilate_i \sont} {\decay} \cdot e^{- \left(\dfrac {\dilate_i \sont} { \decay}-1\right)}  & \text{o/u/shoots} \\
\abs{x_i\left(\dilate\cdot \sont\right)} &x_{i}^\text{max} \geqslant\thresh\\[3mm]
0 & \text{else}.\\
\end{array}\right.
\end{eqnarray}

We applied the ADV model to FRED data setting the sonification parameters $\sJ=5$, $\snu=2$, $\ssigma=1$, $\s{c}_o$, $\s{c}_u$, $\sA$ and $\decay$ as mentioned above and the other parameters as in the ITR model. Fig. \ref{fig:spectrogram_DB1_ADV} shows the spectrogram of the ADV model for user B. Note the additional  harmonic and subharmonic partials for the overshoot and undershoot segments at 0.4, 1.0, 3.2, 3.7 s, and 0.0, 0.7, 3.6 s respectively (audio file 8). As the experienced user A did not produce any excursions beyond the limits, the ADV model yields the same results as the ITR model (see Fig. \ref{fig:spectrogram_DA2_ITR}, audio file 4).
\begin{figure}[ht]
\begin{center}
\includegraphics[width=\linewidth]{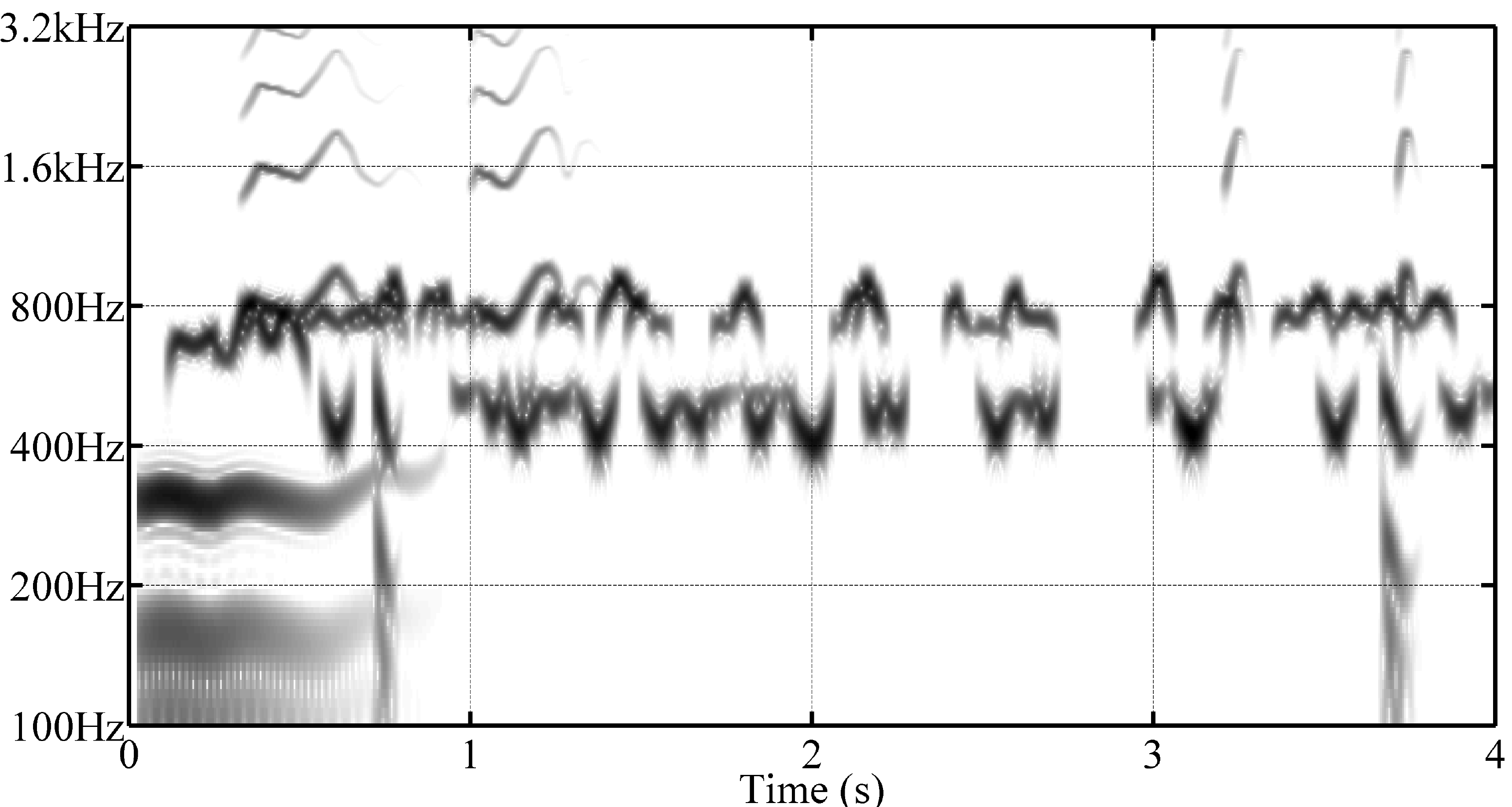}
\caption{Spectrogram of  \DSon\ ADV  model for user B (audio file 8).}
\label{fig:spectrogram_DB1_ADV}
\end{center}
\end{figure}
\section{Conclusion}
The proposed \DSon\ method aims to construct a direct sonification strategy for one-dimensional streams of numerical data. To achieve the intended directness, {\DSon} inherits an important property of other highly direct sonification approaches like audification and auditory graphs, in that it preserves the overall temporal structure of the data stream. {\DSon} is especially well-suited  for data whose size (number of data points), is too small to be suitable for (pure) audification, because the audified sound would be either too short to perceptually decipher data details when using a high playback rate or, otherwise, would be displayed at very low frequencies where the human auditory system lacks good sensitivity. 

H\"oldrich and Vogt's Augmented Audification \cite{Holdrich:2015} addressed the same problem domain. To ameliorate the drawback of the output being in too low a frequency range, they applied a data-dependent single side-band modulation to shift audio up by a desired frequency. The problem with this is that the frequencies in the data are scaled linearly resulting in compression of the frequency relationships, thereby destroying the periodicity of harmonic signals. A solution might be to use pitch-shifting which retains the frequency ratios, but this introduces artefacts into the signal and only works well for small shifts. %

In \DSon's general form, the data stream is cut into non-overlapping segments where the selection of the slicing points depends on the nature of the data and the envisioned application. (In the presented test case of biomechanical data, the zero-crossing points of the trend-free speed signal are utilized as segment boundaries.) Each segment is sonified as a single sonic event using a sonification method not predefined within the general {\DSon} framework. For instance, a method (such as the proposed modified auditory graphs) which is based on mapping data properties of the segment to sound parameters could be used; even a highly metaphorical sonification which displays an alert whenever a segment's duration exceed a certain threshold is possible (though at the cost of reducing directness.) 
To form the entire \DSon\ signal, the sonic events are superimposed in such a way that the temporal pattern of the segments' starting points corresponds precisely to the temporal structure of the cutting points, thereby preserving the overall relative time structure of the data. 

As the sonification method for the segments is structurally decoupled from the formation of the final sound stream, the  playback speed of the entire \DSon\ signal can be set independent of the length of the individual sonic events offering a wide range of possible time compression/stretching factors and thereby high flexibility for zooming into or out of the data. Even pure audification can be regarded as a special case of \DSon, if every single data point is treated as a segment and sonified by a Dirac impulse weighted by the signed data value.

To ensure maximum directness of the resulting sonification, a modified auditory graph has been proposed as the specific method for sonifying the individual segments. In contrast to common auditory graphs, additional amplitude modulation derived from the segment's data evolution in an application-dependent way is accommodated to accentuate large data values. Furthermore, the reference frequency (and thereby the pitch register) is set individually for each sonic event depending on specific segment properties, for example, positive and negative-valued segments in an AC signal, or an overall  trend.

As a demonstration, three \DSon\ models using  variants of modified auditory graphs (with/without AM thresholding and timbre design) were applied to data gathered from FRED exercise sessions. The determination of the cutting points, as well as the specific choice of the amplitude modulation (thresholding in the Individual Target Range model), are based on domain expertise and intended to display the main features of physiotherapeutic interest in a perceptually salient way. For the third advanced model, the modified auditory graph was extended by incorporating a different timbre for segments whose magnitude exceeds a predefined range. 

\DSon\ offers some, albeit limited, potential for real-time applications since a segment's sonic event can generally only be synthesized when its end point is reached and the entire segment is available for deriving parameters of the specific sonification method.

The \DSon\ framework provides a wide range of application-dependent flexibility (as demonstrated by the different models for post hoc analysis of physiotherapeutic data) while maintaining a high degree of directness of the auditory display in that it succeeds in letting the data `speak' for themselves.
For future work, it is intended to apply \DSon\ to data from other domains which allow for the precise determination of specific detection or discrimination tasks, so that the \DSon\ method can be compared with audification and auditory graphs in formal listening tests.
\section*{Acknowledgment}
The authors would like to thank Kirsty Lindsay and Nick Caplan of Northumbria University's Aerospace
Medicine and Rehabilitation Laboratory for their advice on the salient information sought by physiologists in the post hoc analysis of FRED exercise data.
\ifCLASSOPTIONcaptionsoff
  \newpage
\fi
%
%
\bibliographystyle{IEEEtran}
%

\vspace{-12 mm}
\begin{IEEEbiography}[{\includegraphics[width=1in,height=1.24in,clip,keepaspectratio]{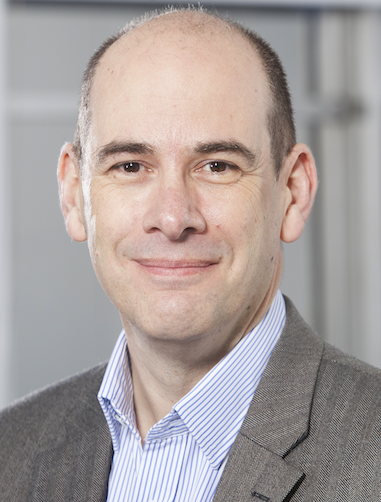}}]
{Paul Vickers}
is Associate Professor in Computer Science at Northumbria University, Newcastle upon Tyne, UK. He is a UK Chartered Engineer, holds a BSc degree in Computer Studies and a PhD in Software Engineering \& HCI. His research focuses on auditory display and sonification. He has served on the board of  the International Community for Auditory Display (2004--2012, 2015--2018). 
\end{IEEEbiography}\vspace{-12mm}
\begin{IEEEbiography}[{\includegraphics[width=1in,height=1.25in,clip,keepaspectratio]{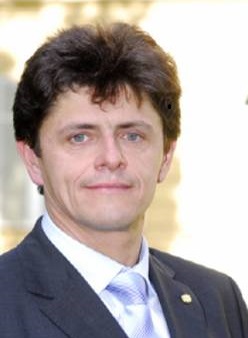}}]
{Robert H\"oldrich} is Full Professor in Audio Engineering and Acoustics at the Institute of Electronic Music and Acoustics (IEM) at the University of Music and Performing Arts Graz, Austria. He studied the flute, composition and electrical engineering and has a doctorate in applied mathematics. His research  includes studies of sonification, psychoacoustics, signal processing, sound analysis and synthesis, recording and playback technology, and virtual acoustics. 
\end{IEEEbiography}

\end{document}